\documentclass[12pt]{article}

\usepackage{amsmath, amsfonts, amssymb}
\usepackage{graphicx}
\usepackage{jheppubcustom}
\newcommand{\de}{\partial}
\def\be{\begin{equation}}
\def\ee{\end{equation}}
\def\beq{\begin{equation}}
\def\eeq{\end{equation}}
\newcommand{\al}{\alpha}
\newcommand{\bt}{\beta}
\newcommand{\gam}{\gamma}

\newcommand{\del}{\delta}

\newcommand{\eps}{\epsilon}

\newcommand{\lam}{\lambda}
\newcommand{\Lam}{\Lambda}

\newcommand{\Sig}{\Sigma}

\newcommand{\om}{\omega}
\newcommand{\Om}{\Omega}

\newcommand{\rmd}{\mathrm{d}}

\newcommand{\Mpl}{M_{\textrm{Pl}}}
\newcommand{\Mp}{M_{\textrm{Pl}}}

\renewcommand{\k}{\vec{k}}

\newcommand{\K}{\mathcal{K}}
\renewcommand{\L}{\mathcal{L}}

\newcommand{\x}{\vec{x}}
\newcommand{\y}{\vec{y}}

\newcommand{\Tr}{\mathrm{Tr}}
\newcommand{\bg}{\bar{g}}
\newcommand{\bSig}{\bar{\Sig}}

\renewcommand{\L}{\mathcal{L}}

\renewcommand{\[}{\left[}
\renewcommand{\]}{\right]}
\renewcommand{\(}{\left(}
\renewcommand{\)}{\right)}

\title{Cosmology and perturbations in massive gravity}

\author{Guido D'Amico}
\affiliation{Center for Cosmology and Particle Physics,\\
Physics Department, New York University, \\
4 Washington Place, New York, NY 10003, USA}
\emailAdd{gda2@nyu.edu}

\toccontinuoustrue

\abstract{We study perturbations around some cosmological backgrounds in the dRGT theory of massive gravity.
We develop a general formalism to calculate the perturbations around any background.
We derive the Lagrangian for fluctuations in the  small scale limit, and for the open FRW solution we repeat the analysis around the full background.
We find that the perturbations display similar properties: the longitudinal modes of the massive graviton are instantaneous at quadratic level, but they acquire a kinetic term at cubic order.
}

\begin{document}

\maketitle

\section{Introduction}

Recently, the possibility of constructing a consistent, Lorentz-invariant theory of massive gravity has received renewed interest.
Historically, the first quadratic Lagrangian which describes a massive spin-2 field was written by Fierz and Pauli \cite{FierzPauli1939}.
Such a Lagrangian propagates 5 degrees of freedom on a Minkowski background, although the mass term (in unitary gauge) breaks all four linearized diffeomorphisms which are a symmetry of the massless theory.
However, Boulware and Deser (BD) \cite{BoulwareDeser1972} showed that the sixth mode, which is a ghost, reappears beyond the quadratic order.
A renewed theoretical interest was spun by the Stueckelberg field formalism introduced in \cite{Siegel1993, ArkaniHamedGeorgiSchwartz2002}, which allows a better understanding of the properties of the theory in terms of the effective theory of the helicity-1 and helicity-0 degrees of freedom.
It becomes apparent that the BD ghost arises from the fact that the helicity-0 field satisfies a fourth-order equation of motion, and its absence in the quadratic Fierz-Pauli Lagrangian is guaranteed by the ``sick'' kinetic term being a total derivative.
One might therefore wonder whether a similar tuning can be found for the non-linear terms in the graviton Lagrangian: the answer is yes, as was recently shown by de Rham and Gabadadze \cite{deRhamGabadadze2010}.
Finally, it was shown by de Rham, Gabadadze and Tolley (dRGT) that it is indeed possible to resum the whole perturbative Lagrangian into a 2-parameter theory, which is fully non-linear and Lorentz-invariant \cite{deRhamGabadadzeTolley2010}.
By construction, such a theory propagates only five modes around Minkowski background, and Hassan and Rosen \cite{HassanRosen2011B} proved that the theory is indeed always free of the sixth mode, as the lapse function remains a Lagrange multiplier after a redefinition of the shift vector.
It is worth stressing that this does not guarantee that the theory is completely free of ghosts.
In fact, around some backgrounds some of the five propagating degrees of freedom might become ghosts or show other instabilities.

In dRGT theory, exact solutions with high degree of symmetry have been found: we refer the reader to \cite{Nieuwenhuizen2011, KoyamaNizTasinato2011, GruzinovMirbabayi2011, DAmicoEtal2011, GumrukcuogluLinMukohyama2011A, ChamseddineVolkov2011, Berezhianietal2011, GratiaHuWyman2012, KobayashiEtal2012} for some spherically symmetric and cosmological solutions.
Some interesting features of these solutions are worth to be pointed out.
For instance, spherically symmetric, static solutions in asymptotically flat space have a physical singularity at the horizon \cite{GruzinovMirbabayi2011}. This is not true if one does not require asymptotic flatness \cite{KoyamaNizTasinato2011, Berezhianietal2011}.
Perhaps, the most striking peculiarity of the dRGT massive gravity is that there are no spatially flat homogeneous and isotropic cosmological solutions \cite{DAmicoEtal2011}: the same constraint that forbids the BD ghost does not allow for a non-trivial cosmological evolution of a flat FRW ansatz.
Allowing for spatial curvature, it was realized in \cite{GumrukcuogluLinMukohyama2011A} that two branches of solutions to the constraint equation open up, both describing an open FRW metric.

An interesting problem to investigate is the behavior of fluctuations around non-trivial backgrounds, because their stability is not guaranteed a priori.
Since the potential of dRGT is written in terms of matrix square roots, a general perturbative expansion presents some technical difficulty.
Up to now, perturbations have been studied in the decoupling limit around Lorentz-invariant backgrounds \cite{deRhamEtal2010}; the issue of superluminality has been considered in \cite{Gruzinov2011, deRhamGabadadzeTolley2011}; linear cosmological perturbations have been studied in \cite{GumrukcuogluLinMukohyama2011B, DeFeliceGumrukcuogluMukohyama2012} and an analysis of perturbations has been performed around the black hole solution in \cite{Berezhianietal2011}.

In this work, we take a small step to understand the behavior of the perturbations around some Lorentz-breaking backgrounds.
First, we develop a general and algorithmic formalism for writing down the fluctuation Lagrangian around generic backgrounds.
We apply the formalism to the three cosmological solutions presented in \cite{KoyamaNizTasinato2011, DAmicoEtal2011, GumrukcuogluLinMukohyama2011A}.
We choose a local inertial coordinate system and we focus on high frequency, high momentum fluctuations in a small region of space-time.
Our results are somewhat surprising.
Around the three backgrounds, the quadratic Lagrangians for perturbations take essentially the same form, and the only propagating modes are the GR ones.
The additional degrees of freedom, in fact, lose their time-kinetic term and their speed of sound is formally infinite.
This degeneracy is broken at cubic order, because interactions induce a time-kinetic term for the additional modes.
For the open FRW solution, we also consider the perturbative Lagrangian around the full non-linear background.
In unitary gauge, we find that the quadratic Lagrangian only propagates the two helicity-2 polarizations as in GR.
At cubic level the degeneracy is broken and the 3 additional modes will propagate.

A similar situation has been encountered in the decoupling limit lagrangian of massive gravity in \cite{deRhamEtal2010, Berezhianietal2011}, in cosmological solutions of the bimetric massive gravity \cite{CrisostomiComelliPilo2012}, and in the non-linear Fierz-Pauli theory in \cite{TasinatoKoyamaNiz2012}.
Also other modified gravity theories, for instance Ho\v{r}ava gravity \cite{BlasPujolasSibiryakov2009}, present similar problems.
Recently, \cite{DeFeliceGumrukcuogluMukohyama2012} showed how the vanishing quadratic time-kinetic term can be obtained as the isotropic limit of perturbations of an anisotropic Bianchi universe.
They also showed that the modes propagate at the quadratic level when a non-vanishing anisotropy is present, and one of them is always a ghost.

At classical level, the kinetic term can have either sign because it is a cubic term.
But even if the square of the frequency has a negative sign, for wavelengths that are not very long the frequency of perturbations would be very large, and therefore above the cutoff of the effective theory.
At the quantum level, a quadratic kinetic term can be generated, as there is no symmetry forbidding it, and its sign will depend on the interactions and parameters of the theory.
However, we expect its coefficient to depend on the cutoff used to regulate the theory.
Another issue to consider is the coupling to matter of such modes, and how do they modify the gravitational interactions among bodies.
These are interesting and deep questions, and we will not address them in this paper, leaving their study for future work.

The paper is organized as follows: in Section \ref{sec:massiveGR} we present the theory.
In Section \ref{sec:perts} we develop the formalism we use to get the fluctuation Lagrangian, and in Section \ref{sec:stability} we explain the physical setup of our analysis.
In Section \ref{sec:solutions} we describe the cosmological backgrounds we are interested in and derive the Lagrangians for perturbations.
Finally, in Section \ref{sec:conclusions}, we discuss our results.


\section{Action of Massive Gravity}
\label{sec:massiveGR}

In general, a Lorentz-invariant action for a theory of massive gravity can be written as
\be
\label{eq:action}
S = \frac{\Mpl^2}{2} \int \rmd^4 x \sqrt{-g} \left[ R - U(g_{\mu \nu}, \phi^a) \right] \; ,
\ee
where $U$ is a potential which depends on the metric $g_{\mu \nu}$ and on four auxiliary scalar fields $\phi^a$, which are introduced in order to keep manifest diffeomorphism invariance \cite{Dubovsky2004}.
The $\phi^a$'s are scalars under diffeomorphisms, but they transform as a Lorentz vector under an internal global $SO(3,1)$ symmetry group, $\phi^a(x) \to \Lam^a_{\; b} \phi^b(x)$.
Matter fields are coupled to $g_{\mu \nu}$ in unitary gauge, defined as $\phi^a = \del_\mu^a x^\mu$: since the matter action is diff-invariant, the scalar fields are not directly coupled to matter.
The most general action of the form \eqref{eq:action} propagates two graviton polarizations plus four other degrees of freedom.
Five of these correspond to the polarizations of a massive graviton, whereas the sixth mode is a ghost, known as the Boulware-Deser ghost \cite{BoulwareDeser1972}.
However, recently there has been a proposal for a theory of massive gravity \cite{deRhamGabadadze2010, deRhamGabadadzeTolley2010}, which has been shown to propagate just the five polarizations of a massive spin-2 field in \cite{HassanRosen2011B, Mirbabayi2011, Kluson:2012wf}  (for a review of massive gravity, see \cite{Hinterbichler2011}).

The non-linear potential presented in \cite{deRhamGabadadzeTolley2010} describes a two-parameter family of theories, and can be written in the form
\be
\label{eq:potential}
U(g_{\mu \nu}, \phi^a) = - m^2 \left[ 2 e_2(\K) + 6 \bt_3  \, e_3(\K) + 24 \bt_4 \, e_4(\K)\right] \; ,
\ee
where $\K$ is the matrix defined by
\be
\K^\mu_{\; \nu} = \del^\mu_{\;\nu} - \sqrt{\Sig}^\mu_{\; \nu} \; ,
\ee
and $\Sig$ is the following combination of the metric and scalar fields:
\be
\Sig^{\mu}_{\; \nu} = g^{\mu \al} \de_\al \phi^a \de_\nu \phi^c \eta_{ac} \, .
\ee
Here $\eta_{ab}$ denotes the Minkowski metric, which is taken as a fixed metric in the target space\footnote{One can modify the theory by considering a generic reference metric, or by considering the second metric to be dynamical as well, in which case the spectrum consists of a massless and a massive graviton \cite{HassanRosen2011C}.}.
We denote by $e_i(\K)$ the following invariants of a $4 \times 4$ metric:
\begin{subequations}
\begin{align}
e_{1}(\K) &= \Tr{\K} \\
e_{2}(\K) &=  \frac{1}{2} \[ (\Tr{\K})^2 - \Tr{\K^2} \]\\
e_{3}(\K) &= \frac{1}{6} \[ (\Tr{\K})^3 - 3 \Tr{\K} \Tr{\K^2} + 2 \Tr{\K^3} \] \\
e_{4}(\K) &= \det{\K} \, ,
\end{align}
\end{subequations}
while $e_k(\K) = 0$  for $k > 4$.

The expression \eqref{eq:potential} written in terms of the matrix $\K$ is useful as $\K$ starts linear in perturbations around the vacuum solution $g_{\mu \nu} = \eta_{\mu \nu}$, $\phi^a = \del^a_\mu x^\mu$.
For our purposes, we will find convenient to rewrite the action directly in terms of the matrix $\Sig^\mu_{\; \nu}$:
\be
\label{eq:action2}
S = \frac{\Mpl^2}{2} \int \rmd^4 x \sqrt{-g} \left\{ R + 2 m^2 \left[ \al_0 + \al_1 e_1(\sqrt{\Sig}) + \al_2 e_2(\sqrt{\Sig})
+ \al_3 e_3(\sqrt{\Sig}) \right] \right\} \, ,
\ee
where the $\al_i$ are not independent, being related to the $\bt_i$ as
\be
\begin{split}
\label{eq:alpha}
&\al_0 = 6 + 12 \bt_3 + 12 \bt_4 \, , \quad \al_1 = - (3 + 9 \bt_3 + 12 \bt_4) \, , \\
&\al_2 = 1 + 6 \bt_3 + 12 \bt_4 \, , \quad ~\ \al_3 = - 3 (\bt_3 + 4 \bt_4) \, .
\end{split}
\ee
Notice that in eq. \eqref{eq:action2} the $e_1$ term appears because we explicitly expand the invariants of $\K$.
Also, we can safely drop the term $e_4(\sqrt{\Sig}) = \sqrt{\det \Sig}$, since it is a total derivative and does not contribute to the equations of motion.

\section{A formalism for perturbations}
\label{sec:perts}

The presence of the square root structure in the action \eqref{eq:action2} makes the analysis of the theory technically difficult.
The problem of writing down the Lagrangian for perturbations around any given background is just a technical one: the expansion of the matrix square roots is difficult because the background matrix $\bSig$ and its perturbations $\del \Sig$ will not commute in general.
One can rely on a Lagrange multiplier formulation of the action \cite{deRhamGabadadzeTolley2010}, or on the formulation in terms of vierbeine \cite{HinterbichlerRosen2012}, but the calculations quickly become cumbersome because of the increased number of fields.
Here we work out algebraic equations which relate the invariants of the matrix $\sqrt{\Sig}$ in terms of the invariants of $\Sig$, which are readily perturbed (our method generalizes the calculations presented in \cite{Gruzinov2011}).
These equations can then be solved order by order to get the Lagrangian for fluctuations.

We start by establishing some notation.
Denoting by $\lam_i$ the eigenvalues of the matrix $\Sig$, by definition the invariants $e_i(\Sig)$ are given by
\begin{subequations}
\begin{align}
& s_1 \equiv e_1(\Sig) = \sum_i \lam_i \, , \\
& s_2 \equiv e_2(\Sig) = \sum_{i<j} \lam_i \lam_j \, , \\
& s_3 \equiv e_3(\Sig) = \sum_{i<j<k} \lam_i \lam_j \lam_k \, , \\
& s_4 \equiv e_4(\Sig) = \lam_1 \lam_2 \lam_3 \lam_4 \, .
\end{align}
\end{subequations}
Therefore, the invariants of the square root $e_i(\sqrt{\Sig})$ have the following expressions:
\begin{subequations}
\begin{align}
& t_1 \equiv e_1(\sqrt{\Sig}) = \sum_i \lam_i^{1/2} \, , \\
& t_2 \equiv e_2(\sqrt{\Sig}) = \sum_{i<j} \lam_i^{1/2} \lam_j^{1/2} \, , \\
& t_3 \equiv e_3(\sqrt{\Sig}) = \sum_{i<j<k} \lam_i^{1/2} \lam_j^{1/2} \lam_k^{1/2} \, , \\
& t_4 \equiv e_4(\sqrt{\Sig}) = \sqrt{\det \Sig} \, .
\end{align}
\end{subequations}
It is straightforward to derive the relations:
\begin{subequations}
\begin{align}
t_1^2 &= s_1 + 2 t_2 \, , \\
t_2^2 &= s_2 - 2 \sqrt{s_4} + 2 t_1 t_3 \, , \\
t_3^2 &= s_3 + 2 t_2 \sqrt{s_4} \, .
\end{align}
\end{subequations}
Finally, one can combine the above identities to get the following quartic equations for the variables $t_1^2$, $t_2$, $t_3^2$:
\begin{subequations}
\begin{align}
\begin{split}
\label{eq:t1}
&t_1^8 - 4 s_1 t_1^6+(6 s_1^2 - 8 s_2 - 48 \sqrt{s_4}) t_1^4 \\
&\phantom{t_1^8 - 4 s_1 t_1^6}
+ 4 (-s_1^3+4 s_1 s_2 - 16 s_3 + 8 s_1 \sqrt{s_4}) t_1^2
+ (s_1^2 -4 s_2 + 8 \sqrt{s_4})^2 = 0 \, ,
\end{split} \\
\label{eq:t2}
&t_2^4 - 2(s_2+6 \sqrt{s_4}) t_2^2 -8 (s_3+s_1 \sqrt{s_4}) t_2
+ (s_2-2 \sqrt{s_4})^2 - 4 s_1 s_3 = 0\, ,  \\
\begin{split}
\label{eq:t3}
&t_3^8 - 4 s_3 t_3^6 + (6 s_3^2 - 8 s_2 s_4 - 48 s_4^{3/2}) t_3^4 \\
&\phantom{t_3^8 - 4 s_3 t_3^6}
+ 4 (-s_3^3+4 s_2 s_3 s_4 - 16 s_1 s_4^2 + 8 s_3 s_4^{3/2}) t_3^2
+ (s_3^2 -4 s_2 s_4 + 8 s_4^{3/2})^2 = 0\, .
\end{split}
\end{align}
\end{subequations}
In principle, these equations can be even solved exactly to express the lagrangian as a scalar function of traces of $\Sig$, but the resulting expression would be quite cumbersome and of very little use.
Instead, we will solve eqs. (\ref{eq:t1}, \ref{eq:t2}, \ref{eq:t3}) perturbatively.
Given an exact background solution $\bg_{\mu \nu}$, $\bar{\phi}^a$, we define the perturbations to the metric and scalar fields as $g_{\mu \nu} = \bg_{\mu \nu} + h_{\mu \nu}$ and $\phi^a = \bar{\phi}^a + \pi^a$\footnote{Another possible definitions for the perturbations of the Stueckelberg is via the formula $\phi^a(t,x^i) = \bar{\phi}^a(t+\pi^0, x^i+\pi^i)$, in which case the $\pi^a$'s have simpler transformation properties under diffeomorphisms.}.
We can construct the background matrix $\bSig^\mu_{\; \nu} = \bg^{\mu \al} \de_\al \bar{\phi}^a \de_\nu \bar{\phi}_a$ and the perturbations $\del \Sig = \Sig - \bSig$.
The invariants will be expanded as $s_i = \bar{s}_i + \del s_i$ and $t_i = \bar{t}_i + \del t_i$, where $\bar{s}_i = e_i(\bSig)$ and $\bar{t}_i = e_i(\sqrt{\bSig})$.
It is straightforward to solve eqs. (\ref{eq:t1}, \ref{eq:t2}, \ref{eq:t3}) for $\del t_i$ order by order as a function of $h_{\mu \nu}$, $\pi^a$.
Such a calculation method is algorithmic and it can be easily performed with a symbolic mathematics program (we used Mathematica in the following calculations).

\section{Small scale expansion}
\label{sec:stability}

The perturbative technique outlined in the previous section is very general and can be used to derive the fluctuation Lagrangian in different physical setups.
It can also be applied to the case in which the reference metric is different from Minkowski or in the bimetric massive gravity theory \cite{HassanRosen2011C}, with minimal modifications.

Here, we want to analyze the degrees of freedom and the stability of the theory around some cosmological backgrounds, by focusing on the Lagrangian of the graviton and Stueckelberg fields in a small region of space-time \cite{Dubovsky2004, DubovskyEtal2005}.

The basic idea is that, for distances much smaller than the curvature radius, we can go to a locally inertial coordinate system such that the (background) metric, around a point $x^\mu = 0$, is a small deformation of flat space-time:
\be
\bg_{\mu \nu}(x) = \eta_{\mu \nu} - \frac{1}{6} (R_{\mu \al \nu \bt} + R_{\mu \bt \nu \al}) x^\al x^\bt + \mathcal{O}(x^3) \, ,
\ee
where $R^\al_{~\bt \mu \nu}$ is the Riemann tensor of the background solution.
In massive gravity, any coordinate transformation will modify the Stueckelbergs as well, so we will expand them in a Taylor series\footnote{The Stueckelberg fields are derivatively coupled, so in principle we should keep the third order in the expansion. However, we are interested to derive the equations of motion for perturbations in the $x \to 0$ limit, in which case the third order does not contribute \cite{Mirbabayi2011}}:
\be
\bar{\phi}^a = A^a_\mu x^\mu + B^a_{~\mu \nu} x^\mu x^\nu + \mathcal{O}(x^3) \, .
\ee
The matrix $A^a_\mu$ is actually the vierbein of the unitary gauge metric at the point $x = 0$ \cite{Mirbabayi2011}, and it will be proportional to $\delta^a_\mu$ only around Minkowski space-time.
In particular, we are spontaneously breaking the local Lorentz invariance even at the point $x = 0$, which will have important consequences at the level of perturbations.
In the next section, we will work out the lagrangian for perturbations around this background, keeping only the terms which survive the $x^\mu \to 0$ limit in the equations of motion.

\section{Cosmological solutions and perturbations}
\label{sec:solutions}

We are interested to study the behavior of perturbations around three background cosmological solutions: the de Sitter solution of ref.  \cite{KoyamaNizTasinato2011}, the flat FRW solution described in the appendix of \cite{DAmicoEtal2011}, and the open FRW solution found in \cite{GumrukcuogluLinMukohyama2011A}.
It has to be noted that the solutions of \cite{KoyamaNizTasinato2011} and \cite{DAmicoEtal2011} are isotropic but inhomogeneous, since it is impossible to find a flat FRW solution in which the scalar field sector is homogeneous and isotropic, as shown in \cite{DAmicoEtal2011}.
However, if the graviton mass is small ($m \ll H$), these solutions approximate very well the GR ones, because of a cosmological Vainshtein mechanism.
If one allows for spatial curvature, an open FRW solution does exist \cite{GumrukcuogluLinMukohyama2011A}, because this is invariant under a $SO(3,1)$ group, which is the diagonal subgroup of the space-time and of the internal $SO(3,1)$'s.

The cosmological solutions admit a coordinate system in which the metric is of the FRW form.
In isotropic coordinates, it can be written as:
\be
\label{eq:FRW}
\rmd s^2 = - \rmd \tau^2 + a^2(\tau) \frac{\rmd \vec y^2}{(1 - |K| \vec y^2/4)^2} \, ,
\ee
where $K \leq 0$ is the curvature constant (we consider only spatially flat or open solutions).
We go to a local inertial frame by performing the following non-linear transformation to new coordinates $t$, $\x$ \cite{NicolisRattazziTrincherini2008}:
\begin{equation}
\label{eq:coords}
\tau = t - \frac{1}{2} H(t) \x^2 \, , \qquad \vec y = \frac{\x}{a(t)} \[ 1 + \frac{H^2(t)}{4} \x^2 \] \, ,
\end{equation}
which brings the metric into the following form (up to second order in $x$):
\be
\rmd s^2 =  - \[ 1 - (\dot{H}(t) + H^2(t)) \x^2 \] \rmd t^2
+ \[ 1 - \frac{1}{2} H^2(t) ( 1 - \Om_K(t)) \x^2 \] \rmd \x^2 \, ,
\ee
where $H = \dot{a}/{a}$ is the Hubble constant and $\Om_K \equiv - K/(a^2 H^2)$ is the usual curvature parameter.
It will be useful to work with coordinates which are a small conformal deformation of Minkowski space-time.
In such a form the metric will deviate from Minkowski both for distances and times $\sim H^{-1}$, and the gauge freedom will be completely fixed.
The needed infinitesimal diffeomorphisms are \cite{NicolisRattazziTrincherini2008}:
\be
\label{eq:xitransf}
\xi_0 = - \frac{1}{4} (2 \dot{H} + H^2 + \Om_K H^2) \(t \x^2 + \frac{1}{3} t^3\) \, ,
\qquad
\xi_i = \frac{1}{4} (2 \dot{H} + H^2 + \Om_K H^2) \x_i t^2 \, ,
\ee
and the metric reads
\be
\rmd s^2 = \[ 1 - \frac{1}{2} (1 - \Om_K) H^2 \x^2 + \frac{1}{2} \( 1 + \Om_K + 2 \frac{\dot{H}}{H^2} \) H^2 t^2 \]
(-\rmd t^2 + \rmd \x^2) \, .
\ee
Notice that in such a form, the parameters $H$, $\dot{H}$ and $\Om_K$ are all evaluated at a fixed time.
The different solutions will differ by the spatial curvature and by the Stueckelberg sectors, which we will now describe in detail.

\subsection{de Sitter solution}

We consider the solution discussed in the main section of \cite{KoyamaNizTasinato2011}, which is restricted to the one-parameter choice $\bt_3 = - 4 \bt_4$.
For notational simplicity, we will denote $C \equiv (1-12 \bt_4)/(1-8 \bt_4)$ in the following formulae.
In FRW coordinates, the metric is de Sitter in planar coordinates, while the Stueckelbergs are given by
\begin{equation}
\phi^0 = \frac{2 C}{3 H \del}
\operatorname{arctanh} \[ \frac{1+ H^2 \rho^2 - e^{-2 H \tau}}{1 - H^2 \rho^2 + e^{-2 H \tau}} \]
- p\( \frac{3}{2 C} e^{H \tau} |\y| \) \, , \qquad \quad 
\phi^i = \frac{3}{2 C} e^{H \tau} y^i \, ,
\end{equation}
%
where $\del$ is a dimensionless integration constant in the range $0 < \del < 4 C^2/9$, and the function $p(z)$ is defined by the equation
\be
p'(z) = - \frac{2}{3 \del} C H z \frac{\sqrt{16 C^4 - 81 \del^2 + 36 C^2 \del^2 H^2 z^2}}{4 C^2 H^2 z^2 - 9} \, .
\ee
Using the transformation to inertial coordinates \eqref{eq:coords} and \eqref{eq:xitransf}, the Stueckelberg background up to second order in $x$ becomes, up to an irrelevant constant,
\begin{equation}
\label{eq:phidS}
\phi^0 =
\frac{2 C}{3 \del} t + \frac{3 H}{4 C} \( \frac{16 C^4}{81 \del^2}-1\)^{1/2} \x^2 \, ,
\qquad \quad
\phi^i = \frac{3}{2 C} x^i \, .
\end{equation}

The quadratic Lagrangian for the perturbations reads:
\be
\label{eq:L2dS}
\begin{split}
\L_2 =& \frac{m^2 \Mp^2}{2} \(1 - 12 \bt_4 \) \( \frac{4 C^2}{9 \del} -1 \) \[ \frac{1}{4} F_{ij}^2
- \frac{3}{2 C} h_{ij} (\de_i \pi_j - \de_k \pi_k \del_{ij})
+ \frac{9}{16 C^2} (h_{ij}^2 - h_{kk}^2) \] + \\
&+ \frac{m^2 \Mp^2}{2} \(1 - 12 \bt_4 \) \sqrt{\frac{4 C^2-9 \del}{4 C^2 + 9 \del}} \[ 4 H \pi^0 \de_j \pi_j - H x^i F_{ij} \de_j \pi^0 - \frac{4 C^2}{9 \del} H x^i F_{ij} \dot{\pi}_j \]
 \, ,
\end{split}
\ee
where we have defined $F_{ij} \equiv \de_i \pi_j - \de_j \pi_i$.

In the high momentum limit, the mixing with gravity can be ignored.
In the Stueckelberg Lagrangian, it is clear that none of the $\pi$'s is propagating, their equations of motion involving only spatial derivatives.
A particularly bad situation arises in the Lorentz-invariant limit, i.e. when the parameter $\del \to 4 C^2/9$, since in this limit the full second-order Lagrangian vanishes: here perturbations are infinitely strongly coupled.

The cubic lagrangian reads:
\begin{multline}
\label{eq:L3dS}
\L_3 = \frac{(1-12 \bt_4) C}{54 \del} \bigg[ (4 C^2 + 9 \del)
F_{ij} F_{jk} \de_{(i} \pi_{k)}
- 36 \del F_{ij} \de_i \pi^0 \dot{\pi}_j \\
+ 9 \del F_{ij}^2 (\dot{\pi}^0 + \de_k \pi_k)
+ 324 \del^2 (\de_{(i} \pi_{j)} - \de_k \pi_k \del_{ij}) (\dot{\pi}_i \dot{\pi}_j + 2 \dot{\pi}_i \de_j \pi^0 + \de_i \pi^0 \de_j \pi^0) 
 \bigg] \, ,
\end{multline}
from which it is clear that there is a time-kinetic term for the $\pi_i$, while $\pi^0$ is algebraically determined.
One might think that this result is an artifact of the choice of gauge, but it is not.
In fact, at quadratic level we wrote down the full Lagrangian for perturbations. Even if we choose a gauge in which the perturbations are only in $h_{\mu \nu}$, it is clear that $h_{00}$ and $h_{0i}$ do not appear and the constraint structure is the same as GR.
At cubic order, we checked that $h_{00}$ appears as a Lagrange multiplier in the mass term, while $h_{0i}$ appear quadratically, thus modifying the GR constraints and giving a kinetic term to three additional modes.
Another doubt one may have is that the cubic order kinetic term could be removed by a field redefinition (we refer the reader to \cite{deRhamGabadadzeTolley2011B} for a discussion on this point).
By writing down the equations of motion, it appears that two initial data are required to specify the solution for $\pi_i$.
Therefore, we conclude that three longitudinal modes are propagating at cubic order.

\subsection{Flat FRW solution}

A generic flat FRW solution has been derived in the Appendix of \cite{DAmicoEtal2011}, for the parameter choice $\beta_3 = \beta_4 = 0$.
The scale factor is determined by the matter content of the Universe, and the Stueckelberg sector is, in FRW coordinates,
\begin{equation}
\phi^0 = \frac{9}{16 T} \int^\tau \frac{\rmd s}{a(s) H(s)} + a(\tau) T \( 1 + \frac{9 \y^2}{16 T^2} \) \, , \qquad \quad
\phi^i = \frac{3}{2} a(\tau) y^i \, ,
\end{equation}
where $T$ is an integration constant with dimensions of time\footnote{This is one particular solution of a non-linear differential equation, every solution of which gives an equally valid Stueckelberg sector for the FRW metric. For instance, taking $\bt_4 = 0$, the dS solution of the previous section obeys such equation.}.
Using the transformation to local inertial frame, up to $\mathcal{O}(H^2 x^2)$, the scalar background becomes
\begin{equation}
\phi^0  = \frac{9+ 16 a^2 H^2 T^2}{16 a H T} t + \frac{16 a^2 H^2 T^2 - 9}{16 a H T}  \[ (H + \dot{H}/H) \frac{t^2}{2}
- H \frac{\x^2}{2} \] \, , \quad
\phi^i = \frac{3}{2} x^i \, .
\end{equation}

The perturbative Lagrangian at quadratic level reads
\be
\label{eq:L2FRW}
\begin{split}
\L_2 =& \frac{m^2 \Mpl}{2} \frac{(3 - 4 a H T )^2}{24 a H T} \[ \frac{1}{4} F_{ij}^2 - \frac{3}{2} h_{ij} (\de_i \pi_j - \del_{ij} \de_k \pi_k)
+ \frac{9}{16} \( h_{ij}^2 - h_{kk}^2 \) \] + \\
&+ \frac{m^2 \Mp^2}{2} \frac{3 - 4 a H T}{3 + 4 a H T} \[ 4 H \pi^0 \de_j \pi_j - H x^i F_{ij} \de_j \pi^0 - \frac{16 a^2 H^2 T^2 + 9}{24 a H T} H x^i F_{ij} \dot{\pi}_j \]
 \, ,
\end{split}
\ee
which has the very same structure of \eqref{eq:L2dS}, and does not propagate any degrees of freedom.

At cubic order, the Lagrangian reads
\begin{multline}
\L_3 = \frac{(3+4 a H T)^2}{144 a H T} F_{ij} F_{jk} \de_{(i} \pi_{k)}
+ \frac{2}{3} F_{ij} \de_i \pi^0 \dot{\pi}_j
- \frac{1}{6} F_{ij}^2 \(\dot{\pi}^0 + \de_k \pi_k \) \\
+ \frac{16 a H T}{(3 + 4 a H T)^2} \(\de_{(i} \pi_{j)} - \de_k \pi_k \del_{ij} \)
\(\dot{\pi}_i \dot{\pi}_j + 2 \dot{\pi}_i \de_j \pi^0 + \de_i \pi^0 \de_j \pi^0 \) 
 \, ,
\end{multline}
which is very similar to \eqref{eq:L3dS}, and the same considerations apply.

\subsection{Open FRW solution}

For the solution presented in \cite{GumrukcuogluLinMukohyama2011A}, the scalar field background reads (in FRW isotropic coordinates)
\be
\phi^0 = c_\pm a(\tau) \sqrt{\frac{1}{|K|} + \frac{\y^2}{(1-|K| \y^2/4)^2}} \, , \qquad \quad
\phi^i = c_\pm a(\tau) \frac{y^i}{1-|K| \y^2/4} \, .
\ee
There are two branches of solutions identified by the constant $c_\pm$:
\be
\label{eq:cpm}
c_\pm \equiv \frac{\gam_1 + \gam_2 \pm \sqrt{\gam_1^2 - \gam_2}}{\gam_2}
\ee
where, to simplify the following expressions, it is useful to define:
\be
\gam_1 \equiv 1+ 3 \bt_3 \, , \qquad \gam_2 \equiv 3 (\bt_3 + 4 \bt_4) \, .
\ee
Since the theory is defined by the positive square roots, we require $c_{\pm} >0$.
In the local inertial frame, up to second order in $H x$, we get
\begin{equation}
\phi^0 = \frac{c_\pm}{\sqrt{\Om_K}} t + \frac{1}{2} \frac{c_\pm}{\sqrt{\Om_K}}
\[ (H + \dot{H}/H) t^2 - (1-\Om_K) H \x^2 \]  \, , \qquad
\phi^i = c_\pm x^i \, .
\end{equation}

As before, we write the quadratic Lagrangian, keeping only the terms which survive the $x \to 0$ limit in the equations of motion:
\be
\label{eq:L2open}
\begin{split}
\L =& \frac{m^2 \Mpl^2}{2} \frac{(1-\sqrt{\Om}_K) A}{\sqrt{\Om}_K}
\[ \frac{1}{4 c_\pm} F_{ij}^2
- h_{ij} \(\de_i \pi_j - \de_k \pi_k \del_{ij}\)
+ \frac{c_\pm}{4} (h_{ij}^2 - h_{kk}^2) \] + \\
&+ \frac{m^2 \Mp^2}{2} \frac{(1-\sqrt{\Om_K}) A}{c_\pm}
 \[ - 4 H \pi^0 \de_j \pi_j + H x^i F_{ij} \de_j \pi^0 + \frac{1}{\sqrt{\Om_K}} H x^i F_{ij} \dot{\pi}_j \]
 \, ,
\end{split}
\ee
where we have defined $A = (1 + 2 \gam_1 + \gam_2 - c_\pm (\gam_1+\gam_2) )$.
Again, this Lagrangian has a similar structure to \eqref{eq:L2dS} and \eqref{eq:L2FRW} and describes just non-propagating modes.
This result resonates with the conclusions of \cite{GumrukcuogluLinMukohyama2011B}, who analyzed the cosmological perturbations at linear order and found instantaneous propagation of the vector and scalar modes.
We will extend their analysis around the fully non-linear background in the next section.

An interesting observation can be made regarding the $\Om_K \to 1$ limit.
In fact, the choice $\Om_K = 1$ corresponds to the Milne universe solution, which has the metric $\rmd s^2 = - \rmd \tau^2 + \tau^2 \rmd \x^2$ in FRW coordinates, but it is just a reparametrization of Minkowski space-time (in fact, the metric becomes exact Minkowski in local coordinates).
The quadratic Lagrangian \eqref{eq:L2open} in this limit turns out to be $\L = 0$, which signals that the perturbations become infinitely strongly coupled.
This has to be compared with the healthy lagrangian $\L = - \frac{1}{4} F_{\mu \nu}^2 + \dots$ around the Minkowski solution.
The difference in behavior should not surprise us.
In fact, the scalar equations of motion for the open FRW ansatz have three branches of solutions \cite{GumrukcuogluLinMukohyama2011A}: two are present only in the case $K \neq 0$ and correspond to the non-trivial cosmological evolution we just described, and the third one forbids any evolution of the scale factor, so that the metric reduces to flat space-time.
So, perturbing around the ``cosmological'' branch and then taking the Minkowski limit does not commute with considering the Minkowski branch and then perturbing around it: the first choice displays a bad behavior.

At cubic order, the Lagrangian reads
\be
\label{eq:open_cubic}
\begin{split}
\L_3 &= \frac{1}{4 c_\pm \sqrt{\Om}_K}
\[  ( \gam_1 + \gam_2 - 3 c_\pm \gam_2)  F_{ij} F_{jk} \de_{(k} \pi_{i)} - c_\pm \gam_2 F_{ij}^2 \de_k \pi_k \] \\
&- \frac{\gam_2}{24}
\[ 48 \( \de_{(i} \pi_{j)} -\de_k \pi_k \del_{ij} \) \dot{\pi}_i \de_j \pi^0
- 3 \( G_{ij}^2 - G_{kk}^2 \) \dot{\pi}^0
+ 2 G_{ij} G_{jk} G_{ki} - 3 G_{ij}^2 G_{kk} + G_{kk}^2 \] \\
&- \frac{\gam_1 + \gam_2 - 2 c_\pm \gam_2}{c_\pm} \( F_{ij} \de_i \pi^0 \dot{\pi}_j - \frac{1}{4} F_{ij}^2 \dot{\pi}^0 \) \\
&+ \frac{\gam_1 + \gam_2 - 2 c_\pm \gam_2}{c_\pm} \sqrt{\Om_K}
\(\de_{(i} \pi_{j)} - \de_k \pi_k \del_{ij} \)
\(\dot{\pi}_i \dot{\pi}_j + 2 \dot{\pi}_i \de_j \pi^0 + \de_i \pi^0 \de_j \pi^0 \) 
 \, ,
\end{split}
\ee
where $G_{ij} \equiv 2 \de_{(i} \pi_{j)}$, and for simplicity we performed an expansion in the small parameter $\Om_K$, retaining the leading order for each term.
Again, the $\pi^0$ is algebraically determined while the $\pi_i$ acquire a time-kinetic term.

\subsection{Cosmological perturbations}

The analysis of perturbations on small scales highlights the degeneracy of the cosmological backgrounds.
To convince ourselves that this is not an artifact of the approximations made, it is also worth considering the behavior of perturbations also around the fully non-linear background solutions, without any small-scale expansion.
This is easy to do in the case of the open FRW solution, which we will now study in detail.

We start by writing the background in the following coordinate system:
\be
\rmd s^2 = - \rmd t^2 + a^2(t) \Om_{ij} \rmd x^i \rmd x^j
= - \rmd t^2 + \frac{a^2(t)}{1+ |K| \x^2} (\x^2 \del_{ij} - x^i x^j) \rmd x^i \rmd x^j \, ,
\ee
where, $K < 0$ is the value of the spatial curvature.
The background Stueckelbergs are of the form
\be
\bar{\phi}^0 = \frac{c_\pm a(t)}{\sqrt{|K|}} \sqrt{1 + |K| \x^2} \, , \qquad \bar{\phi}^i = c_\pm a(t) x^i \, .
\ee
We then choose the unitary gauge for the perturbations, i.e. we consider $g_{\mu \nu} = g^{FRW}_{\mu \nu} + h_{\mu \nu}$ and we leave the $\phi^a$'s unperturbed.
Applying the method of section \ref{sec:perts}, we find the mass term of the Lagrangian at quadratic order to be \cite{GumrukcuogluLinMukohyama2011B}
\begin{multline}
\L_m^{(2)} = \L_{c.c.} + \frac{\Mpl^2}{8} m^2 a^3 \sqrt{\det \Om_{ij}} \, \frac{1- \sqrt{\Om_K}}{\Om_K} \\
\times c_\pm \[ (\gam_1+\gam_2) c_\pm - 3 (1+2 \gam_1+\gam_2) \] \(h^2 - h^{ij} h_{ij}\)
\end{multline}
where we have defined $h \equiv \Om^{ij} h_{ij}$, $h^{ij} \equiv \Om^{ik} \Om^{jl} h_{kl}$.
The term $\L_{c.c.}$ is the effective cosmological term $- \Mpl^2 \Lam_{\rm eff} \sqrt{-g}$ expanded at second order, where
\be
\Lam_{\rm eff} = - \frac{m^2}{\gam_2} \[ 2 c_\pm (\gam_1^2-\gam_2) - (\gam_1 +2 \gam_1^2 - 2 \gam_2) \] \, .
\ee
From this expression it is clear that in the full Lagrangian only two degrees of freedom are propagating, as in GR.
In fact, $h_{00}$ and $h_{0i}$ do not show up at all in the mass term, so that the full Lagrangian has the same constraint structure as the GR one.
We can now ask if this enhanced symmetry is respected by the cubic order Lagrangian, and the answer is no.
In fact, the expression for the mass term at cubic order reads
\be
\begin{split}
\L_3 &= \frac{\Mpl^2 m^2 a^3 \sqrt{\det \Om}}{16 \gam_2 \sqrt{\Om_K}} \bigg[
\( \frac{c_\pm}{3} (6 \gam_1^2 + \gam_2^2 -5 \gam_2 + 2 \gam_1 \gam_2)
- (\gam_1+\gam_2) (1+2 \gam_1+\gam_2) \) h_{ij} h^{jk} h_k^{\;i} \\
&+ \( (\gam_1+\gam_2) (1+2 \gam_1+\gam_2) - 2 c_\pm (\gam_1^2-\gam_2)\)  h_{ij} h^{ij} h - \frac{c_\pm}{3} \gam_2 (1+3 \gam_1+\gam_2) h^3 \\
&+ \sqrt{\Om_K} \( (\gam_1+\gam_2) (1+2 \gam_1+\gam_2) - c_\pm (2 \gam_1^2 + 2\gam_1\gam_2 - \gam_2 +\gam_2^2) \)
h_{00} (h^{ij} h_{ij} - h^2) \\
&+ 4 \Om_K  \( (\gam_1+\gam_2) (1+2 \gam_1+\gam_2) - c_\pm (2 \gam_1^2 + 2\gam_1\gam_2 - \gam_2 +\gam_2^2) \)  (h^{ij} - h \Om^{ij}) h_{0i} h_{0j} \bigg] \, ,
\end{split}
\ee
where, for simplicity, we only display the leading terms in an expansion in $\Om_K$ for each coefficient.
Looking at this expression, we see that, while $h_{00}$ appears as a Lagrange multiplier, $h_{0i}$ appears quadratically, which means that the cubic Lagrangian propagates three more modes than the quadratic one.
This is in fact analogous to what we have seen in the small scale limit\footnote{Notice that the term involving $h_{0i} h_{0j}$ appears with a coefficient proportional to $\sqrt{\Om_K}$, as the term involving $\dot{\pi}_i \dot{\pi}_j$ in eq. \eqref{eq:open_cubic}. This is a consistency check for our calculations.}.

\section{Conclusions}
\label{sec:conclusions}

The cosmological solutions we have considered look different from each other.
However, they share similar properties at the level of perturbations.
Namely, longitudinal modes of the graviton do not propagate at quadratic order in perturbations, but they do get a time-kinetic term by cubic order interactions.

This means that, schematically, their dispersion relation is of the form $\k^2 = 0$ at quadratic level, but it is modified to $\k^2 = \eps \om^2$ when considering the cubic Lagrangian.
Here $\eps$ is actually the fluctuation field, whose sign in particular is ambiguous.
Therefore, perturbations around the cosmological background can have negative frequency squared, but for short wavelength the frequency is actually very high, and above the cutoff of the effective theory.
Quantum effects can generate a quadratic kinetic term, whose sign will depend on the parameters and interactions of the theory. We expect that its coefficient will depend on the cutoff used to regulate the theory.
An important question to address is to include the coupling to matter, and to understand the experimental consequences of the behavior of the longitudinal modes.
In general, it would be interesting to understand what is the reason for the vanishing of the kinetic terms, and in particular how this behavior is related to the symmetries of the auxiliary and physical metric.

All these questions are important and interesting, and we leave them for future studies.

\section*{Acknowledgements}
It is a pleasure to thank Lasha Berezhiani, Paolo Creminelli, Sergei Dubovksy, Gregory Gabadadze, Nemanja Kaloper, Matt Kleban Mehrdad Mirbabayi, Shinji Mukohyama, Lorenzo Sorbo for useful and stimulating discussions.
I would like to acknowledge the Aspen Center for Physics and the NSF Grant \#1066293 for hospitality during the final stages of this work.
The author is supported by a James Arthur Fellowship.

\bibliography{Bibliography}

\end{document}